\documentclass[intlimits,twoside,a4paper]{article}

\usepackage{amsmath,amssymb}
\usepackage{graphicx}
\usepackage[T2A]{fontenc}

\usepackage[cp1251]{inputenc}

\usepackage{cmpj2}

\issue{2013}{16}{4}{43003}
\doinumber{10.5488/CMP.16.43003}

\title[Time evolution of a small reactive system]{Time evolution of a small reactive system\thanks{I am very happy to dedicate this work to my old friend and great scientist
Professor M. Holovko for the celebration of his 70th birthday.}}
\author[J.P. Badiali]{J.P. Badiali}
\address{LECIME, ENSCP-Universit\'e Pierre et Marie Curie, CNRS/UMR7575
4 Place Jussieu, 75230 Paris Cedex 05, France}

\date{Received July 15, 2013, in final form August 25, 2013}

\begin{document}

\maketitle

\begin{abstract}
We investigate the irreversible evolution of a small system in which a chemical reaction
takes place. We have two main goals: the first requires to find an equation
to produce a time-irreversible behavior,the second consists in introducing a simple
exactly solvable model in order to understand basic facts in chemical kinetics.
Our basic tool is the transition function counting the number of paths joining
two points in the reactive coordinates system. An exact quantum Smoluchowski
equation is derived for the reactive system in vacuum, in the presence of a solvent
in equilibrium at any time with the reactive system a new Smoluchowski equation is obtained.
The transition from a quantum regime to a classical one is discussed.
The case of a reactive system not in equilibrium with its neighborhood is investigated
in terms of path integral and via a partial differential function. Memory effects
and closure assumptions are discussed. Using a simple potential model, the chemical
rate constant is exactly calculated and questions such as the meaning of the
activation energy or the physical content of the so-called prefactor are investigated.

\keywords irreversibility, transition function, Smoluchowski equation,
chemical rate constant, activation energy
\pacs 03.65.Ca, 05.30.-d, 05.70.-a, 47.53.+n
\end{abstract}

\section{Introduction}
The transformation of matter via chemical reactions is a traditional and fundamental subject
of investigations in Chemical Physics and in Statistical Physics. Due to the existence of new
experimental methods and to the increase in the capacity of computers the domain is renewed.
As concerns the theoretical investigations of the chemical rate constant there is a clear
distinction between modelling approaches in which we search to gain an insight on the possible
theoretical interpretation and the simulation approaches that we use to compare theory
and experiments. This paper pertains to the first category, we do not discuss the second one
which is essentially based on the solution of a time-dependent Schr\"{o}dinger equation
(see for instance the feature article \cite{koslov}).

Hereafter we want to discuss very fundamental questions such as: what is the role of dynamics in
the activation energy usually considered as quasi-thermodynamics quantity, how we may observe
the transition from a quantum to a classical regime, what happens if the reactive system is
not in equilibrium with a solvation shell. In order to do that a simple model will be investigated.
Some results have been already given in \cite{jpb1} and \cite{jpb2} and some technical points already presented will not
be repeated herein where we focus on a new presentation thereof and on new results in presence of a solvation shell.
This paper is organized as follows. In section~2 we give a very brief overview concerning
the traditional approaches of the chemical kinetics. In section~3 we develop the basic ingredients
of our approach in the case of a reactive system in vacuum. In that case a quantum Smoluchowski
equation is exactly derived. In section~4, these results are extended in order to describe a
chemical reaction in which the reactive species are in thermal equilibrium with a solvent
at any time. In section~5 we analyze the effects on the chemical reaction of a small number of particles
forming a solvation shell not in equilibrium with the reactive system.
These effects are discussed in the path integral formalism or via a partial differential equation.
In section~6 we discuss exact results concerning the chemical rate constant, in particular, the
meaning of the activation energy and the physical content of the prefactor are analyzed.
Conclusions are presented in section~7.

\section{A short overview on the subject}
A first important step in the description of the chemical kinetics has been proposed by
Arrhenius in $1910$ \cite{laidler} for whom the chemical rate constant $k$ is related to the
activation energy $E$, the temperature $T$ and the so-called pre-factor $A$ according to
\begin{equation}
k = A \exp\left\{-\frac{E}{k_{\mathrm{B}}T}\right\}\,.
\label{arrhenius}
\end{equation}
Frequently used to fit experimental data, this relation does not give a clear physical meaning
both to $E$ and $A$.

In $1935$ the transition state theory (TST) proposed by Eyring \cite{eyring} represented a
second major step in the approach of chemical kinetics.
In the Eyring's formulation, it is established that
\begin{equation}
k = K \frac{k_{\mathrm{B}}T}{\hbar} \exp\left\{-\frac{\Delta G}{k_{\mathrm{B}}T}\right\}
\label{eyring}
\end{equation}
in which $K$ is a transmission coefficient and $\Delta G$ is a Gibbs free energy.
The TST is at the origin of a huge literature and it is not our aim to discuss it among
a lot of papers, books, review articles published on the subject. We can mention a review
paper dedicated to H. Eyring published in $1983$ \cite{trular1} and a
more recent review paper published in $1996$ \cite{trular2}.

The fundamental assumptions on which the TST is built up are as follows: (i) there exists
a surface separating the reactant region from the product region and no recrossing of this
surface is allowed, (ii) there is an equilibrium for the reactants implying also an equilibrium
between reactants and products. This last point made of the TST a quasi-thermodynamic approach.
In its initial formulation, the TST is basically a time independent theory describing
what happens in gas phase. Later on some extensions to describe reactions in solution were introduced and it appears that some quantities assumed to have a quasi thermodynamic
origin may contain a large part of dynamic processes. The presence of a solvent may strongly
affect the chemical rate constant and the problem concerning the equilibrium or the
non-equilibrium of the solvation shell is crucial.

To describe the irreversible behavior of chemical reactions, in $1940$ Kramers \cite{kramers}
used a Fokker-Planck equation as a starting point for calculating the chemical rate constant.
Today this stochastic approach is no more disconnected from the TST when we consider chemical
reactions in a liquid state where we have to predict a transition from the TST to a diffusion controlled regime.
In the Kramers work, the Fokker-Planck equation results from very general aspects in the theory of
random processes. Today there is a large literature from which this equation is derived
from a physical basis. The system+reservoir methods have been introduced to derive such an equation
in quantum physics (for a review in this field see \cite{weiss}).
In these approaches, the system under investigation is considered as a small open part of a large system and
the dissipation arises from the energy transfer from the small system to its large environment.
In the absence of reservoir, the system is described by the Schr\"{o}dinger equation and
the processes are time-reversible.
In the presence of a reservoir, the dynamics in a small system is described by a quantum mechanical
Langevin equation \cite{grabert} or equivalently by a generalized quantum master equation that
is frequently a Fokker-Planck type equation \cite{caldeira}. Thus, the effect of the reservoir is
to change the dynamics of a small system.
Therefore, it does not correspond to the traditional concept of
thermostat used in standard statistical mechanics.

The system+reservoir approach has proved to be a very efficient method of treating the problems such
as the effect of dissipation processes on the quantum tunneling. More recently,
in relation to molecular electronics, it appears very important to understand the coupling
of individual molecular structures with a bath of bosons under non-equilibrium
conditions \cite{ratner,jun}. We have to deal with new questions such as the definition of a local temperature (see for instance \cite{galperin}) or the problem of heat
dissipation in molecular junctions (see for example \cite{pecchia}).
In that case, a non-equilibrium Green function formalism has been introduced from the
work by Datta \cite{datta}. This approach leads to a kinetic equation that can be reduced
to a diffusion equation in particular conditions \cite{datta}.
However, there exist small systems for which we observe an irreversible behavior in the absence
of reservoir; trivial examples are the reaction like $A + B \to AB$ in a dilute gas
phase or the nuclear fission processes considered in the seminal paper by Kramers \cite{kramers}.

In what follows our reactive system corresponds to a particle $A$ inserted in a small box
in which some particles noted globally $B$ create an external fixed potential. This potential
can result from quantum mechanical calculations, which is defined in reactive coordinates
space (see for instance \cite{sacha}). When an equilibrium state is reached, $A$ will
be attached in one or several parts of $B$ for which the potential exhibits a minimum.
Thus, we mimic the reaction $A + B \to AB$. This reactive system will be put in contact
with different kinds of environment.

We decide to characterize the evolution of this system by a transition function defined
in terms of the paths in reaction coordinates. We do not claim that it is the only possibility
to describe the motion, but without doubt it is a possible choice. Hereafter we want
to analyze the consequences of this choice. We first consider the reactive system isolated in vacuum.

\section{A reactive system in vacuum}

The transition function is a real valued function that describes the transition in space-time
from the point $x_{0}$ at the initial time $t_{0}$; to the point $x$ at the time $t$ it is given
by a functional integral according to
\begin{equation}
q(x_{0},t_{0}; x,t) = \int \mathcal{D}x(t) \exp\left\{- \frac{1}{\hbar} A[x_{0},t_{0}; x,t]\right\}
\label{transition}
\end{equation}
in which $\mathcal{D}x(t)$ is the measure for the functional
integral and $A[x_{0},t_{0}; x,t]$ is the euclidean action
\begin{equation}
A[x_{0},t_{0}; x,t] = \int\limits_{t_{0}}^{t} \left\{\frac{1}{2} M \left[\frac{\rd x(s)}{\rd s}\right]^{2} + u[x(s)]\right\} \rd s
\label{euclidaction}
\end{equation}
in which $M$ is the mass of the particle and $u[x(s)]$ is the external potential; hence, each path
is weighted by the total energy spent on this path.
If the points $x_{0}$ are distributed in space according to $\phi_{0}(x_{0})$, we may introduce
a function $\phi(t,x)$ defined as
\begin{equation}
\phi(t,x) = \int \rd x_{0} \phi_{0}(x_{0}) q(x_{0},t_{0}; x,t).
\label{deffi}
\end{equation}
The time evolution of $\phi(t,x)$ is given by the following partial
differential equation \cite{jpb2}
\begin{equation}
-\frac{\partial  \phi(t,x)}{ \partial t} + D\Delta_{x}
\phi(t,x) -\frac{1}{\hbar}{u[x(t)] \phi(t,x)} = 0
\label{dif}
\end{equation}
in which $\Delta_{x}$ means the laplacian operator taken at the point $x$ and $D = \frac{\hbar}{2M}$
is a quantum diffusion coefficient. This equation has been previously derived in (\cite{jpb2}) and the technical points
in the derivation will not be reproduced here. In addition, a derivation in more general conditions is given in the subsection $5.2$
of this paper.

At this point, we must underline several points. The solution of (\ref{dif}) cannot be normalized
and consequently $\phi(t,x)$ is not a probability. The transition function defined
via (\ref{transition}) is a real valued function defined in terms of real quantities and
(\ref{dif}) is not a Schr\"{o}dinger equation, and of course $\phi(t,x)$ is not a wave function and it has a physical
meaning by itself. Nevertheless, $\phi(t,x)$ gives a quantum description of the system since due to the functional
integral the paths have a fractal character, there are energy fluctuations along the paths
and these fluctuations are measured relatively to $\hbar$ \cite{jpb4}. In comparison with what
is done in system+reservoir approaches \cite{grabert}, we do not start with the Schr\"{o}dinger
equation. Our basic quantity verifies a time irreversible equation having a quantum origin
and it exists in the absence of a solvent. According to our approach, it will be possible
to describe an irreversible chemical reaction in vacuum.

Instead of $\phi(t,x)$, as proposed in \cite{kampen}, we may introduce a
function $P(x,t) =  \phi(t,x) \exp\left\{V/2\right\}$ in which $V(x)$ is a dimensionless potential that
can be calculated as it is shown below. It is easy to show that $P(x,t)$ verifies the Smoluchowski equation
\begin{equation}
\frac{\partial P(t,x)}{\partial t} =
D \Delta_{x}P(t,x) + D \nabla_{x}\{[\nabla_{x}V(x)]P(t,x)\}
\label{smol}
\end{equation}
and we may conclude that contrary to $\phi(t,x)$, the function $P(t,x)$ is a probability.
Note that the dimensionless potential $V(x)$ is different from $\beta u(x)$ usually expected
(for instance in \cite{kramers}).

\section{A reactive system in the presence of a solvent}

For brevity the reactive particle $A$ will be hereafter referred to as the ``Particle''.
The action  associated with the Particle has been given in \eqref{euclidaction}.
This Particle is coupled with a bath of $N$ identical particles. The center of the spatial
coordinates will be the center of the small box in which the Particle is inserted and we
assume to have a one dimensional system to save the notations. The bath is represented
by $N$ independent oscillators (see \cite{jpb2} for more details). The energy associated with
the bath is given by a second order perturbation around the initial positions, and we have
\begin{equation}
U[R^{N}(t)] = U[R^{N}(t_{0})] + \sum_{i} \frac{1}{2} m \omega^{2} r_{i}^{2}(t)
\label{potbath}
\end{equation}
in which $U[R^{N}(t_{0)}]$ is the bath energy at the initial
time $t_{0}$, $r_{i}$ means for $i$ its derivation from its initial position,
and $\omega^{2}= {\frac{1}{m} \frac{\partial^{2}U[R^{N}(t_{0})]}{\partial R_{i}(t_{0})^{2}}}$.
The Hamiltonian corresponding to the bath is
\begin{equation}
H_{\mathrm{B}}(t)= \sum_{i} \left\{\frac{1}{2}m \left[\frac{\rd r_{i}(t)}{\rd t}\right]^{2} +  \frac{1}{2} m \omega^{2} r_{i}^{2}(t)\right\}.
\label{hbath}
\end{equation}
This model for the bath is commonly used \cite{caldeira83}.

For the interaction between the bath and the Particle we use a Hamiltonian frequently
retained in the literature \cite{caldeira,grabert,caldeira83},
\begin{equation}
H_{\mathrm{C}}(t) = \sum_{i} \mathrm{C} r_{i}(t) x(t)
\label{hinter}
\end{equation}
in which $C$ is the coupling constant.

Instead of $q(x_{0},t_{0}; x,t)$ we now  have to introduce a transition function
$q[x_{0},r^{N}(0),t_{0}; x,r^{N}, t]$ depending on the positions of the $N$ particles of the bath.
This transition function is a generalization of \eqref{transition} defined according to
\begin{equation}
q\left[x_{0},r^{N}(0),t_{0}; x,r^{N}, t\right] = \int \mathcal{D}x(t) \exp\left\{- \frac{1}{\hbar} A[x_{0},t_{0}; x,t]\right\}
\int \mathcal{D}r^{N}(t) \exp\left\{- \frac{1}{\hbar} \int\limits_{t_{0}}^{t}\left[H_{\mathrm{B}}(s) + H_{\mathrm{C}}(s)\right]\rd s\right\},
\label{newq1}
\end{equation}
where $A[x_{0},t_{0}; x,t]$ is given by (\ref{euclidaction}).
In \eqref{newq1}, $r^{N}(t)$ represents the set of deviations from the initial positions
as defined above, $r^{N}(t) = [r_{1}(t), \dots  r_{i}(t), \dots r_{N}(t)]$.
Due to the quadratic form of $[H_{\mathrm{B}}(s) + H_{\mathrm{C}}(s)]$ the functional integral on the
variables $r^{N}(t)$ can be exactly calculated using a mathematical trick
introduced in \cite{feynman2}. We write $r_{i}(t) = r_{i}(t)_{\mathrm{opt}} + \delta r_{i}(t)$
in which $r_{i}(t)_{\mathrm{opt}}$ corresponds to the optimization of the euclidean action associated
with $[H_{\mathrm{B}}(s) + H_{\mathrm{C}}(s)]$. Here, the optimization is a mathematical procedure leading to
a differential equation which is not the equation of motion, in contrast with what is
done in pure quantum mechanics where the optimization of the Lagrangian action is performed. We can write
\begin{equation}
\int \mathcal{D}r^{N}(t) \exp\left\{- \frac{1}{\hbar}\int\limits_{t_{0}}^{t}\left[H_{\mathrm{B}}(s) + H_{\mathrm{C}}(s)\right]\rd s\right\} =
C_{\delta} \exp\left\{- \frac{1}{\hbar} \int\limits_{t_{0}}^{t}\left[H_{\mathrm{B}}(s) + H_{\mathrm{C}}(s)\right]\rd s\right\}_{\mathrm{opt}}
\label{newq2}
\end{equation}
in which the subscript $opt$ means that we have to calculate the trajectories on the
optimum paths and $C_{\delta}$ is a quantity independent of the particle positions, which
results from the integration on the variable $\delta r_{i}(t)$. For two arbitrary
times $t$ and $t_{1} > t$ for which the end positions
correspond to $(x,r^{N})$ and $(x_{1},r^{N}_{1})$ respectively, we define
\begin{equation}
\delta A\left[x,r^{N},t; x_{1},r^{N}_{1},t_{1}\right] = \left\{\int\limits_{t}^{t_{1}} \left[H_{\mathrm{B}}(s) + H_{\mathrm{C}}(s)\right]\rd s\right\}_{\mathrm{opt}}.
\label{deltaA}
\end{equation}
Using the Hamiltonians (\ref{hbath}) and (\ref{hinter}) we can get the explicit
form of $\delta A[x,r^{N},t; x_{1},r^{N}_{1},t_{1}]$ by changing $\omega$ into an
imaginary quantity $i \omega$ in the results given in (\cite{grabert} page $126$, equation $3.23$),
we have
\begin{eqnarray}
\lefteqn{\delta A\left[x,r^{N},t; x_{1},r^{N}_{1},t_{1}\right]  =  \sum_{i}\Bigg\{\frac{1}{2}\frac{m \omega}{\sinh{\omega(t_{1} - t)}}\left[(r_{i,1}^{2}+ r_{i}^{2})\cosh{\omega(t_{1} - t)} - 2 r_{i} r_{i,1}\right] }\qquad\qquad\qquad\nonumber\\
&&{} + \frac{C r_{i}}{\sinh{\omega(t_{1} - t)}} \int\limits_{t}^{t_{1}} \rd s x(s) \sinh{\omega(t_{1}-s)} +
\frac{C r_{i,1}}{\sinh{\omega(t_{1} - t)}}\int\limits_{t}^{t_{1}} \rd s x(s) \sinh{\omega(s-t)} \nonumber\\
&&{} - \frac{C^{2}}{m \omega\sinh{\omega(t_{1} - t)}}\int\limits_{t}^{t_{1}}\rd s x(s) \int\limits_{t}^{s}\rd u x(u)\sinh{\omega(t_{1}-s)}\sinh{\omega(s-t)}\Bigg\}.
\label{hdt}
\end{eqnarray}
Now we can write
\begin{equation}
q\left[x_{0},r^{N}(0),t_{0}; x,r^{N}, t\right] = C_{\delta} \int \mathcal{D}x(t) \exp\left\{- \frac{1}{\hbar}\left\{A\left[x_{0},t_{0}; x,t\right] + \delta A\left[x_{0},r^{N}(0),t_{0}; x,r^{N},t\right]\right\}\right\}
\label{newq4}
\end{equation}
and the transition function for the small system $\bar q(x_{0},t_{0}; x, t)$ can be written
\begin{equation}
\bar q(x_{0},t_{0}; x, t)= C_{\delta} \int \mathcal{D}x(t) \exp\left\{- \frac{1}{\hbar}A_{\mathrm{P}}[x_{0},t_{0}; x,t]\right\}
\left\langle\exp\left\{- \frac{1}{\hbar}\delta A\left[x_{0},r^{N}(0),t_{0}; x,r^{N},t\right]\right\}\right\rangle_{\mathrm{bath}}
\label{qbar}
\end{equation}
in which $\langle\dots \rangle_{\mathrm{bath}}$ means that we have to take a procedure in order to eliminate
the bath particle positions. This procedure depends on the physics under consideration.

We may observe that (\ref{qbar}) is very different and simpler than the expression of
similar quantities calculated in system+reservoir approaches \cite{caldeira,grabert}; this is due
to the fact that we do not start from the density matrix for the overall system but
from the transition function and consequently we do not have to deal with the influence functional.
Another basic difference stands from the fact that the small system when isolated from the bath
is not described by a Schr\"{o}dinger equation but by the time irreversible equation~(\ref{dif}).

The bath we consider is formed of a very large number of particles uniformly distributed in a volume
infinitely large in comparison with the box size. We assume that the bath is in thermal equilibrium
in the field created by the Particle for any value of $t$, a hypothesis commonly accepted
when we study the coupling between a small system and a bath (see for instance \cite{datta}).
Now we must conciliate such a description with the fact
that $\delta A[x_{0},r^{N}(0),t_{0}; x,r^{N},t]$ is defined by an integral as shown by
(\ref{deltaA}), e.g., we must give a meaning to this integral and decide
how to use (\ref{hdt}). This has been done in \cite{jpb2}. The solvent equilibrium at each
time introduces the temperature via a characteristic time $\tau = \hbar/k_{\mathrm{B}}T$.
Hereafter, we focus on time intervals much larger than $\tau$, a common assumption when we
try to derive a Smoluchowski or a Fokker-Planck equation; it is well known that such equations
do not describe short time processes. The final result obtained in \cite{jpb2} shows that the solvent creates a friction
on the reactive particle and the diffusion coefficient $D = \hbar/2M$ is replaced by
\begin{equation}
\bar{D} = \frac{\hbar}{2(M + \mu)}
\label{deff}
\end{equation}
in which $\mu$ is an effective mass, and accordingly the potential $V(x)$ must be
changed in $\bar{V}(x)$. Note that the result (\ref{deff}) is not surprising, because it shows that the coupling of a small
system with a large external system leads to the introduction of an effective mass, which a very well accepted idea in Solid State Physics.
The interest of the derivation given in \cite{jpb2} is to determine the precise conditions leading to(\ref{deff}).
The transition function $\phi(t,x)$ is the solution of the following equation
\begin{equation}
-\frac{\partial  \phi(t,x)}{ \partial t} + \bar{D}\Delta_{x}
\phi(t,x) -\frac{1}{\hbar}{u[x(t)] \phi(t,x)} = 0
\label{newdif_}
\end{equation}
and the corresponding Smoluchowski equation is given by
\begin{equation}
\frac{\partial P(t,x)}{\partial t} =
\bar{D} \Delta_{x}P(t,x) + \bar{D} \nabla_{x}\left\{\left[\nabla_{x}\bar{V}(x)\right]P(t,x)\right\}.
\label{newsmol}
\end{equation}
From (\ref{deff}) we can go from the quantum case $(M \gg  \mu)$ to the classical one ($\mu \gg  M$).

\section{Non-equilibrium between a reactive system and its solvation shell}

In the previous sections we have shown that from a transition function it is possible
to derive a well known equation in statistical physics. This equation is exact in vacuum
and it has been derived under very acceptable conditions in the presence of the solvent that is
assumed to be in equilibrium at any time with the reactive system. However, if it was of interest
to derive such a Smoluchowski equation, the equilibrium condition does not
correspond to the most general situation. The reactive medium is frequently surrounded by
a small number of particles that are not in equilibrium with it. They form a solvation shell
of the reactive system. It is generally accepted that the solvation shell is composed
of a few tens of particles. Hereafter we consider such a case and we try to investigate the
change in chemical reaction due to this solvation shell. This is a standard problem in
chemical kinetics \cite{trular2}.

As above, we describe the surrounding of the reactive system as a set of $N$ oscillators
having the same frequency $\omega$ and the same coupling constant $C$. We first analyze
how the function $\phi(t,x)$ is modified and then we derive a partial differential
equation for this function.

\subsection{Expression of $\phi(t,x)$ in the
limit $\omega(t - t_{0}) \to \infty $}

Let us start from (\ref{hdt}). It is straightforward to analytically perform the integration
on the variables $r^{N}_{1}$ since we have a Gaussian form. The integration over $r^{N}$ requires
to introduce the density of probability on these initial variables. In order to get a tractable
result this density of probability is assumed to have a Dirac form, which means
that $R^{N}(t_{0})$ represents a set of equilibrium positions
and that $r_{i}(t_{0}) =0$. Moreover, we focus on the
limit $\omega(t - t_{0}) \to \infty $, and by this limit we focus on a time interval $t - t_{0}$ much larger
than the period $\frac{2 \pi}{\omega}$ of the bath oscillators. It is well known that equations such as Smoluchowski
or Fokker-Planck are restricted to long time behavior and this limit provides a quantitative meaning to long time.
The final result can be written as follows:
\begin{equation}
\int \rd r^{N}_{1}\rd r^{N} \exp\left\{ -\frac{1}{\hbar} \delta A\left[x,r^{N},t; x_{1},r^{N}_{1},t_{1}\right]\right\} = \bar{C} \exp\left\{-\frac{1}{\hbar} \delta A(x_{0},t_{0}; x,t)\right\}
\label{eq1}
\end{equation}
in which the $\bar{C} = \left[\frac{2 \pi \hbar }{m \omega}\right]^{\frac{1}{2}}$
and $\exp\left\{-\frac{1}{\hbar} \delta A(x_{0},t_{0}; x,t)\right\}$ is given by
\begin{equation}
\exp\left\{\frac{C^{2}}{\hbar m \omega} \int\limits_{t_{0}}^{t}\rd s x(s) \frac{\sinh\omega(t-s)}{\sinh\omega(t- t_{0}}\right\} \int\limits_{t_{0}}^{s}\rd u x(u)
\left[\frac{\sinh(t-u)}{\sinh\omega(t- t_{0})} + \sinh \omega(u -t_{0})\right].
\label{eq2}
\end{equation}
In the limit $\omega(t - t_{0}) \to \infty $
we have $\frac{\sinh\omega(t-s)}{\sinh\omega(t- t_{0}} \approx \exp\left\{- \omega(s- t_{0})\right\}$ and
$\frac{\sinh\omega(t-u)}{\sinh\omega(t- t_{0}} \approx \exp\left\{- \omega(u- t_{0})\right\}$ and we can
rewrite (\ref{eq2}) as
\begin{equation}
\exp\left\{\frac{C^{2}}{\hbar m \omega} \int\limits_{t_{0}}^{t}\rd s x(s) \exp{- \omega(s- t_{0})} \int\limits_{t_{0}}^{s}\rd u x(u) \cosh\omega(u - t_{0})\right\}.
\label{eq3}
\end{equation}
Thus, the existence of a solvation shell that is in non-equilibrium with the reactive system drastically
changes the nature of the problem since a memory effect is introduced.
This result is expected because R. Zwanzig \cite{zwanzig} has shown, by a projectors technics,
that when we want to decouple a small system from a large one, there appears  a memory effect.

However, simple approximations can be introduced in (\ref{eq3}). Quantities such
as $x(s)$ or $x(u)$ have a value limited by the size of the box while the
function  $\cosh \omega(u - t_{0})$ has no restricted value in the limit we consider. In the last integral
of (\ref{eq3}), the integrand is maximum when $u=s$ and we
approximate $\int\limits_{t_{0}}^{s}\rd u x(u) \cosh\omega(u - t_{0})$ by
$x(s) \int\limits_{t_{0}}^{s}\rd u \cosh\omega(u - t_{0}) = \frac{x(s)}{\omega} \sinh\omega(s - t_{0})$.
This is equivalent to correct the external potential $u[x(t)]$ by an
additional term $\delta U[x(t)]$ in the functional integral. In the limit $\omega(t - t_{0}) \to \infty $ we have
\begin{equation}
\delta U[x(t)] = - \frac{C^{2}Nx(t)^{2}}{2 m \omega^{2}} \left[ 1 - \exp\left\{-2 \omega(t- t_{0})\right\}\right] \approx - \frac{C^{2}Nx(t)^{2}}{2 m \omega^{2}}\,.
\label{pot}
\end{equation}
This term will be discussed in the next section.

\subsection{A differential equation}

To derive a differential equation for $\phi(t,x)$ we have to generalize the
calculations presented in \cite{jpb2} leading to (\ref{dif}). Let us consider
\begin{equation}
\phi\left(t + \epsilon, x,r^{N}\right) = \int \phi\left(t,x', r'^{N}\right) q\left(x',r'^{N},t; x,r^{N}, t +\epsilon\right)\rd x' \rd r'^{N}.
\label{newfi}
\end{equation}
The l.h.s. of this equation can be expanded as follows:
\begin{equation}
\phi\left(t +\epsilon,x,r^{N}\right) = \phi\left(t,x,r^{N}\right) + \epsilon \frac{\partial\phi(t,x,r^{N}}{\partial t} + \cdots.
\label{lhs}
\end{equation}
In the r.h.s. of (\ref{newfi}), we first expand $\phi(t,x', r'^{N})$ near $\phi\left(t,x,r^{N}\right)$ and we have
\begin{eqnarray}
\phi\left(t,x', r'^{N}\right)  &=& \phi\left(t,x, r^{N}\right) + \left(x'-x\right)\frac{\partial \phi\left(t,x, r^{N}\right)}{\partial x} + \frac{1}{2}\left(x'-x\right)^{2}\frac{\partial^{2}\phi\left(t,x,r^{N}\right)}{\partial^{2}x} + \sum_{i}\left(r_{i}'-r_{i}\right)\frac{\partial \phi\left(t,x, r^{N}\right)}{\partial r_{i}} \nonumber\\
&&{}
+ \frac{1}{2} \sum_{i}\left(r_{i}'-r_{i}\right)^{2}\frac{\partial^{2}\phi\left(t,x, r^{N}\right)}{\partial^{2}r_{i}^{2}} +\sum_{i}\left(x' - x\right)\left(r'_{i} - r_{i}\right)\frac{\partial^{2}\phi\left(t,x, r^{N}\right)}{\partial x \partial r_{i}} + \cdots .
\label{rhs1}
\end{eqnarray}
To expand $q(x',r'^{N},t; x,r^{N}, t +\epsilon)$ in terms
of $\epsilon$, we consider (\ref{euclidaction}) and (\ref{hdt}), and they become
\begin{equation}
A[x',t; x, t + \epsilon]= \frac{1}{2}\frac{M(x' - x)^{2}}{\epsilon} + \epsilon u(x) +\cdots ,
\label{euclidactionnew}
\end{equation}
\begin{equation}
\delta A[x',r'^{N};x,r^{N},t+\epsilon] = \sum_{i} \frac{1}{2}\frac{m(r'_{i} - r_{i})^{2}}{\epsilon}  + \epsilon C x r_{i}  + \cdots .
\label{hdtnew}
\end{equation}
As expected, $\exp\left\{- \frac{1}{\hbar}\{A[x',t; x, t + \epsilon] + \delta A[x',r'^{N};x,r^{N},t+\epsilon]\}\right\}$
contains a product of Gaussian
functions $G(x'-x) =  \exp\left\{-\frac{m}{2 \hbar \epsilon}(x'- x)^{2}\right\}$ and $G(r'_{i}-r_{i}) = \exp\left\{-\frac{m}{2\hbar \epsilon}(r'_{i}-r_{i})^{2}\right\}$ corresponding to the free motion determined by the kinetic energy.
Now we have
\begin{equation}
q\left(x',r'^{N},t; x,r^{N}, t +\epsilon\right) = \mathcal{N} G(x'-x)\prod_{i}G(r'_{i}-r_{i})\left\{1 - \frac{\epsilon}{\hbar} \left[u(x) + C\sum_{j} x r_{j}\right] + \mathcal{O}(\epsilon^{2})\right\}
\label{newq}
\end{equation}
in which $\mathcal{N}$ is a normalization constant. Due to the Gaussian quadratures, all the
linear terms in $(x'-x)$ or in $(r'_{i}-r_{i})$ vanish, and the quadratic terms
such as $(x' -x)^{2}$ or $(r'_{i}-r_{i})^{2}$ lead to the results linear in $\epsilon$. Thus, in the
expansion we have only to consider a small number of terms. The selection of terms of zero order
in $\epsilon$ determines the constant $\mathcal{N}$ while the linear terms lead
to the following differential equation
\begin{equation}
\frac{\partial  \phi\left(t,x,r^{N}\right)}{ \partial t} = \frac{\hbar}{2M}\Delta_{x}\phi\left(t,x,r^{N}\right) + \frac{\hbar}{2m}\sum_{i}\Delta_{r_{i}}\phi\left(t,x,r^{N}\right)
 -\frac{1}{\hbar}{u(x) \phi\left(t,x,r^{N}\right)} - \frac{C}{\hbar}\sum_{i} x r_{i} \phi\left(t,x,r^{N}\right).
\label{newdif}
\end{equation}
To focus on the properties of the reactive system, we consider the function $\phi(t,x)$ defined according to
\begin{equation}
\phi(t,x) = \int \phi\left(t,x,r^{N}\right) \rd r^{N}\,.
\label{fiinteg}
\end{equation}
Note that the bath is first eliminated because we perform an integration from its initial variables requiring
to introduce a density of probability for them (a Dirac distribution is used), and secondly because the final variables
are irrelevant. In contrast with the case of a reactive system in equilibrium with a large system
where the temperature of the bath is a well defined quantity, in this subsection the solvation
shell is a small system for which the temperature
is not defined and it is not present here. Of course, in the future it should be of interest to investigate a
small system in contact with a solvation shell and immersed in a bath. In this case, the temperature of the bath will appear.
From (\ref{fiinteg}) and (\ref{newdif}) we get the following differential equation
\begin{equation}
-\frac{\partial  \phi(t,x)}{ \partial t} + D\Delta_{x}
\phi(t,x) -\frac{1}{\hbar}{u[x(t)] \phi(t,x)} = - \frac{\hbar}{2m}\int \sum_{i}\frac{\partial^{2} \phi\left(t,x,r^{N}\right)}{\partial r_{i}^{2}}\rd r_{i} + \frac{C x}{\hbar}\int \sum_{i}r_{i}\phi\left(x,r^{N},t\right)\rd r_{i}\,.
\label{difnew}
\end{equation}
The l.h.s. of this equation is identical to (\ref{dif}) while the r.h.s. represents
the coupling with the solvation layer. Due to this coupling, the equation is not closed on
the variables of the reactive system and to get $\phi(t,x)$ we have to know $\phi\left(t,x,r^{N}\right)$.
Thus, as in the previous subsection we see that the existence of a non-equilibrium solvation shell
changes the nature of the problem. In one case we have a memory effect while here we have to deal with the equation
that is not closed on the reactive system variables.
Then,  approximations can be introduced as in the previous subsection. Still we decide to
work on the limit $\omega(t - t_{0}) \to \infty $ and restrict the calculations
to the order $C^{2}$. The details concerning the calculations are reported in
appendix~A and the final result is as follows:
\begin{equation}
-\frac{\partial  \phi(t,x)}{ \partial t} + D\Delta_{x}\phi(t,x) -\frac{1}{\hbar}{u[x(t)] \phi(t,x)} = - \frac{1}{4}\frac{C^{2} x^{2}(t)}{\hbar m \omega^{2}} \phi(t,x)
\label{result}
\end{equation}
Thus, the approximations are sufficient to close the partial differential on the reactive system
variables. Our result is equivalent to the introduction of a correction to $u[x(t)]$ given by
\begin{equation}
\delta U[x(t)] = - \frac{1}{4}\frac{C^{2}N x^{2}(t)}{ m \omega^{2}}\,.
\label{newpot}
\end{equation}
The results (\ref{pot}) and (\ref{newpot}) that are obtained with different approximations
differ just by a numerical factor $(1/2)$.

\section{The chemical rate constant}

We have established three differential equations for $\phi(t,x)$. In vacuum we have (\ref{dif}), and
in the presence of a solvent, the diffusion coefficient changes according to (\ref{deff}) and due
to a solvation layer the potential $u[x(t)]$ should be corrected by $\delta U[x(t)]$.
Three similar Smoluchowski equations are associated to these differential equations.

In order to illustrate these previous results, we study the dynamics of a particle injected in a
small box in which $u(x)$ is a fixed symmetric double well potential. This model has been
subject of many experimental and theoretical investigations in view of a better understanding
of elementary chemical reactions \cite{sacha}. Hereafter we will not repeat all the calculation
details given in \cite{jpb1} and \cite{jpb2}. In order to simplify the notations we work
in a one dimensional system.

The potential $u(x)$ is located in the interval $-b \leqslant x \leqslant b$,
at $x= \pm b$, we put an infinite repulsive barrier. Inside the box,
in the region $ -a \leqslant x \leqslant a$ with $ a < b $, there is a repulsive barrier
of height $U_{1}$ while in
the remaining intervals $[-b,-a]$ and $[a,b]$ there exists an attractive
potential of magnitude $-U_{0} (U_{0} >0)$.
We assume that the Particle of mass $M$ is inserted, at the time $t=t_{0}$ at
the point $x_{0}= -(b+a)/2$, the state
associated with this particle has a spatial extension, $\sigma$, assumed to be
very small in comparison with $(b-a)/{2}$.
Due to this, we can say that the reactant is entirely located in $x <0$ at the initial time.

\subsection{Effect of a solvation shell in non-equilibrium with the reactive system}

The presence of a solvation shell induces both qualitative and quantitative effects on
the chemical kinetics. First, in principle, memory effects have to be introduced. However, from
simple approximations we have shown that the presence of a non-equilibrium solvation
shell may lead to a local correction of the external potential. It is given
by $\delta U[x(t)] = - \lambda \frac{1}{4}\frac{C^{2} x^{2}(t) N}{ m \omega^{2}}$, where the
parameter $\lambda$ may vary from $1$ to $2$ depending on the approximations used to
forget the memory effects or to close the differential equation [see (\ref{pot}) and (\ref{newpot})].
The quadratic dependence of $\delta U(x)$ on $x$ might suggest that the presence of
oscillators in the solvation shell is transmitted via a vibrational effect to the reactive particle .
This naive interpretation is totally misleading since we have seen from (\ref{difnew}) that $\delta U(x)$
results from the coupling potential and from the existence of a laplacian relative to the
coordinates $r_{i}$. As expected, $\delta U[x(t)]$ depends on the number of particles in the
solvation layer, $N$, and it is proportional to $C$, with our assumption that the order $C^{2}$ is retained.
This potential is also inversely proportional to the mass $m$ and to the frequency of the vibrators $\omega$.

The second change due to the solvation layer is quantitative. The quantity $\delta U[x(t)]$
is a negative correction to $u(x)$ showing the possibility to cancel the double well
potential and to replace it by a parabola having a maximum at $x = 0$. In the absence of more information, in what follows
we assume that $\delta U(x)$ does not drastically change the shape of $u(x)$.

\subsection{Solution of the Smoluchowski equation}

The previous results can be summarized by saying that the transition function is the solution
of the following equation:
\begin{equation}
-\frac{\partial  \phi(t,x)}{ \partial t} + \tilde{D}\Delta_{x}\phi(t,x) -\frac{1}{\hbar}\tilde{u}[x(t)] \phi(t,x)
= 0
\label{sintfi}
\end{equation}
in which $\tilde{D}$ can be given by $\frac{\hbar}{2M}$ or
by (\ref{deff}) and $\tilde{u}(x)$ can be $u(x)$ or $u(x) - \delta U(x)$. A Smoluchowski equation corresponds to (\ref{sintfi})
\begin{equation}
\frac{\partial P(t,x)}{\partial t} =
\tilde{D} \Delta_{x}P(t,x) + \tilde{D} \nabla_{x}\left\{\left[\nabla_{x}\tilde{V}(x)\right]P(t,x)\right\}.
\label{sintsmol}
\end{equation}
The solutions of (\ref{sintsmol}) can be expanded based on the solutions of functions $\phi_{n}(t,x)$
 of (\ref{sintfi}). We can write $\phi_{n}(t,x) = f_{n}(t) \varphi_{n}(x)$ in which
$f_{n}(t)= \exp[-({E_{n}- E_{0}}){\hbar^{-1}t}]$, where $E_{0}$ is the energy of the
fundamental state and $E_{n} -E_{0}$ is an eigenvalue of the equation
\begin{equation}
\tilde{D} \Delta_{x}
\varphi_{n}(x) + \frac{1}{\hbar}\left[(E_{n}- E_{0}) - \tilde{u(x)}\right]\varphi_{n}(x) = 0.
\label{phin}
\end{equation}
In vacuum, (\ref{phin}) is exactly a stationary Schr\"{o}dinger equation for a
particle in an external potential $u(x)$, in the presence of a solvation layer, $u(x)$ is
corrected by $\delta U(x)$ while in the presence of a solvent, we have to use (\ref{deff}). The
functions $f_{n}(t)$ are monotonously decreasing functions of time in contrast with the
solutions of the time-dependent Schr\"{o}dinger equation that are
oscillatory functions of time. Using the closure relation between the eigenfunctions we
can write the fundamental solution of (\ref{sintsmol}) as follows:
\begin{equation}
P(0,y;t,x) = \varphi_{0}(x) \sum^{\infty}_{n=0}\left[\frac{\varphi_{n}(y)}{\varphi_{0}(y)}\right] \varphi_{n}(x)\exp\left(-\frac{E_{n}- E_{0}}{\hbar}t\right),
\label{solfond}
\end{equation}
where $P(0,y;0,x) = \delta(x-y)$ as a consequence of the closure relation.
If $f(y)$ is the initial distribution, we define
\begin{equation}
P(t,x) = \int P(0,y;t,x) f(y) \rd y.
\label{sol}
\end{equation}
From (\ref{solfond}) we see that in the limit $t \rightarrow \infty$, the
function $P(0,y ;t,x)$ is only determined by the fundamental state,
$P(0,y ;t,x)$ becomes independent of $y$, and provided that $f(y)$ is normalized
we get $P(0,y;\infty,x) = P(\infty,x) = P_{\mathrm{eq}}(x)= \varphi_{0}(x)^{2}$. The solution of (\ref{sintsmol})
for the long time behavior is also $P_{\mathrm{eq}}(x) = c \exp - \bar{V}(x)$ leading to (see \cite{jpb2} for details)
\begin{equation}
\tilde{V}(x) = -2 \ln \varphi_{0}(x)
\label{defV}
\end{equation}
showing that our system of equations is closed.
In agreement with \cite{kramers,kampen,jpb2},  we interpret
$P_{\mathrm{eq}}(x)$ as the equilibrium density of probability being at the point $x$ where there is an effective
potential $\tilde{V}(x)$. More generally, we can rewrite $P(t,x)$ as follows:
\begin{equation}
P(t,x) = \left[\varphi_{0}(x)\right]^{2}\sum ^{\infty}_{n=0}c_{n}\left[\frac{\phi_{n}(x,t)}{\phi_{0}(x)}\right] = P_{\mathrm{eq}}(x) \gamma(t,x),
\label{newpxt}
\end{equation}
where
\begin{equation}c_{n}= \int \frac{\varphi_{n}(y)}{\varphi_{0}(y)}f(y) \rd y.
\label{cn}
\end{equation}
Thus, before reaching an equilibrium state, $P(t,x)$ is the product of a Boltzmann distribution
by a quantity $\gamma(t,x)$ associated with the paths.

\subsection{The exact chemical rate constant for a simple model}

The first important result of our approach is that the relevant potential is no more the external
potential $u(x)$ given in reaction coordinates. Instead of $u(x)$ we have to deal
with $\tilde{V}(x)$ given by (\ref{defV}). In vacuum, $\tilde{V}(x)$ is a strict transformation
of $u(x)$ related to the fundamental solution of stationary Schr\"{o}dinger equation. In this case,
at least by an example, it has been shown \cite{jpb2} that the shape of $V(x)$ is reminiscent of
the one of $u(x)$. In contact with a solvation shell, $u(x)$ is first modified
by $\delta U(x)$ and dynamic effects such as the mass or the frequency of particles
in the solvation layer are introduced into $\tilde{V(x)}$.
In the presence of a solvent, $\tilde{V}(x)$ depends on the diffusion
coefficient $\bar{D}$ of the reactive particle that is, at least partially, determined
by the friction induced by the solvent. Thus, in the case of vacuum, the activation
energy is expected to be a non-thermodynamic quantity.

The chemical rate constant is defined according to
\begin{equation}
k(t) = - \frac{1}{\mathcal{P}(t)} \frac{\rd\mathcal{P}(t)}{\rd t}
\label{rate1}
\end{equation}
in which
\begin{equation}
\mathcal{P}(t) = \int^{0}_{-b}  P(t,x) \rd x  = \int^{0}_{-b} \phi_{0}(x) \sum ^{\infty}_{0} c_{n} \phi_{n}(x,t)\rd x.
\label{p1}
\end{equation}
The derivative $\frac{\rd\mathcal{P}(t)}{\rd t}$ calculated directly from (\ref{p1}) is given by
\begin{equation}
\frac{\rd\mathcal{P}(t)}{\rd t} = \int^{0}_{-b} \phi_{0}(x) \sum ^{\infty}_{0} c_{n} \frac {\partial \phi_{n}(x,t)}{\partial t}\rd x =
 - \int^{0}_{-b} \phi_{0}(x) \sum ^{\infty}_{0} c_{n} \left(\frac{E_{n} - E_{0}}{\hbar}\right)\phi_{n}(x,t)\rd x.
\label{dpdt1}
\end{equation}
Using the equation of evolution of $\phi_{n}(x,t)$ and after performing an integration by parts, we get
\begin{equation}
\frac{\rd\mathcal{P}(t)}{\rd t} = \left\{\frac{\partial}{\partial x} \bar{D} \left[\phi_{0}(x) \Sigma c_{n} \phi_{n}(x,t)\right]\right\}_{x=0}.
\label{dpdt2}
\end{equation}
We can interpret this as the derivative of the flux crossing the surface at $x=0$.
By simple calculations, it is possible to show that at the initial time the rate constant is given by
\begin{equation}
\left[\frac{\rd\mathcal{P}(t)}{\rd t}\right]_{t=0} =
\bar{D} \left[\frac{\partial f(x)}{\partial x}\right]_{x=0}\,.
\label{ktzero}
\end{equation}
Since we have assumed that the initial distribution $f(x)$ is entirely localized on the
left hand part of $x=0$, we have $k(0)= 0$. Thus, the probability of crossing the dividing surface vanishes at $t=0$ showing that the
rate constant is a time-dependent quantity, as expected.

We can investigate the behavior of $k(t)$ for very large values of $t$. In that case the time-dependent part
of $\mathcal{P}(t)$ is restricted
to $\mathcal{P}_{1}(t)= \smallint \rd x P_{\mathrm{eq}}(x) c_{1}{[\phi_{1}(x,t)]}/{[\phi_{0}(x)]}$; in the same
condition, the calculation
of ${\rd\mathcal{P}(t)}/{\rd t}$ via (\ref{dpdt1}) is trivial. This leads to a stationary
chemical rate constant given by
\begin{equation}
k_{\mathrm{sta}} = \frac{E_{1} - E_{0}}{\hbar}\,.
\label{ksta1}
\end{equation}
From the values of $(E_{1} - E_{0})$ given in \cite{kampen} and the value of $\tilde{V}(0)$, we obtain
\begin{equation}
k_{\mathrm{sta}} =   2\pi \bar{D} \Omega_{0} \omega_{0} \exp\left\{- \tilde{V}(0)\right\},
\label{rate2}
\end{equation}
where the spatial frequencies $(\Omega_{0}, \omega_{0})$ are defined according
to \cite{kramers}. This result looks like the one of Kramers \cite{kramers}. However, our result is exact in
vacuum where we have the dimensionless potential $\bar{V}(0)$ instead of the
quantity $\beta u(x=0) = \beta U_{1}$ that enters the Kramers formula. Moreover, in vacuum there
is no fitting parameter.

In the presence of a solvent at thermal equilibrium, the temperature is fixed and a comparison with
other approaches of the chemical rate constant is more evident. In \cite{jpb2} it has been shown that
\begin{equation}
\tilde{V}(0) = \alpha \left(\frac{U_{1}}{k_{\mathrm{B}}T}\right) = \frac{1}{k_{\mathrm{B}}T}\left[8 U_{1} U_{\xi}\right]^{\frac{1}{2}}
\end{equation}
in which $\alpha =\left[8\frac{1}{U_{1}} \frac{\xi\ a^{2}}{\tau}\right]^{\frac{1}{2}}$ , $\xi = {(\beta \bar{D})^{-1}}$
is a friction coefficient
and $U_{\xi} = {\xi\ a^{2}}{\tau^{-1}}$ represents the energy dissipation associated with the
friction when the Particle crosses the repulsive region with a mean velocity ${a}/{\tau}$.
If the Kramers result reproduces the Arrhenius law with a temperature-independent activation energy, here
there is no reason to assume that $\alpha$ is temperature independent.

In (\ref{rate2}) the prefactor can be rewritten as follows:
\begin{equation}
\left(\frac{k_{\mathrm{B}}T}{\hbar}\right)\left\{2\pi\left[\frac{2 \bar{D} \tau}{(b-a)^{2}}\right]\right\}.
\label{prefactor}
\end{equation}
We recover the traditional factor $({k_{\mathrm{B}}T}{\hbar^{-1}})$ existing in the transition state theory.
It is multiplied by
$[{2 D\tau}{(b-a)^{-2}}]$ that represents the ratio between the mean square
length $\lambda^{2} = 2 D \tau$ associated with a diffusion process occurring during
a time $\tau$ and the square of the thickness $(b-a)$ of the attractive region. Finally, the stationary
rate constant can be written:
\begin{equation}
k = \left(\frac{k_{\mathrm{B}}T}{\hbar}\right) \left[2 \pi (\frac{\lambda}{b-a})^{2}\right] \exp\left\{-\left( \alpha \frac{U_{1}}{k_{\mathrm{B}}T}\right)\right\}.
\label{rate}
\end{equation}
Even if our approach exhibits some similarities with the transition state theory, still the two approaches
are different on several essential points \cite{trular2}: the paths we consider may recross the
dividing surface, we do not use an equilibrium conditions between the products and reactants and we
are able to calculate the time dependence of the chemical rate.

\section{Conclusions}

The results obtained in this paper can be considered at two different levels.
From a statistical mechanical point of view we have shown that it is sufficient to count
the number of paths in classical space-time to derive an equation which is time irreversible
provided that each path is weighted by the total energy spent along the path.
This has been first done in vacuum where exact results are obtained.
A quantum Smoluchowski equation is derived. In the presence of a solvent in equilibrium
with a small system we may again obtain a Smoluchowski equation and a transition
from a quantum regime to a classical one may be observed. However, this is not
the general situation since a system may be not in equilibrium with its surroundings.
This has been investigated in the case of a solvation layer. Then, a memory effect
or a non-closed differential equation appear. With simple assumptions it is possible
to show that the solvation shell leads to an additional potential but local in time.
Although we do not use the Schr\"{o}dinger equation, a stationary Schr\"{o}dinger-like
equation appears and concepts such as ground state or excited states remain relevant.

Our approach is quite different from the system+reservoir methods since we are able to
describe the irreversibility for a small system in vacuum. We do not say that our approach
is the only one possible but it is probably the simplest one. Starting from a real
valued function, i.e., the transition function, we have explored the properties of this function.

Of course, a question may appear: is this formalism capable of describing  time-reversible systems
or systems at thermal equilibrium? This point has been discussed in several
papers \cite{jpb5,jpb6,jpb7,jpb8}. To recover the usual
thermodynamics, we have introduced an entropy function depending on the time at which
the paths are explored, this time is determined by imposing that the mean value of the
energy spent on the paths is equal to the free energy needed to create the system.
From this equilibrium condition we recover all the exact expressions of thermodynamic
quantities. In terms of processes, in order to have reversibility, we have to add
a second equation concerning the reverse system to $\phi(t,x)$ and, using the results obtained
in \cite{nagasawa}, a Schr\"{o}dinger equation can be derived.

Our approach has been used to investigate the time evolution of a reactive system
enclosed in a small box. We have shown that the potential $u(x)$ obtained in principle
by quantum chemistry methods should be replaced by $V(x)$. In vacuum, $V(x)$ is a simple
mathematical transformation of $u(x)$, while in all other situations $V(x)$ includes dynamic
properties such as the frequency of the oscillators forming the solvation shell or a
friction coefficient due to the solvent. When the reactive system is in thermal equilibrium
with a solvent, a temperature appears and we may put our result in a traditional form.
In general the activation energy is no more a thermodynamic quantity. When the energy
associated with the potential $V(x)$ and the frictional energy have the same order
of magnitude, we recover the Arrhenius law. In parallel, the so-called prefactor has been analyzed.

In the future, the role of a solvation shell will be analyzed more profoundly since it can
qualitatively change the nature of the problem by introducing memory effects as well as
quantitatively, since as it has been shown that it may destroy the possibility of a chemical reaction.

\appendix
\section{Calculation of the r.h.s. of (\ref{difnew})}
In order to calculate the first term in the r.h.s of (\ref{difnew}) we can
remark that $\phi\left(t,x,r^{N}\right)$ can be factorized in functions $\phi(t, x, r_{i})$ since
the oscillators are assumed to be independent, and we have
\begin{equation}
I = \frac{\hbar}{2m}\int \sum_{i}\frac{\partial^{2} \phi\left(t,x,r^{N}\right)}{\partial r_{i}^{2}}\rd r_{i} = \frac{\hbar}{2m}N \phi(t,x) \frac{\int\frac{\partial^{2} \phi(t,x,r_{i})}{\partial r_{i}^{2}}\rd r_{i}}{\int \phi(t,x, r_{i}) \rd r_{i}}
\end{equation}
and in the limit $\omega (t - t_{0}) \to \infty$, we have
\begin{equation}
\phi(t,x, r_{i}) \approx \exp\left\{- \frac{1}{\hbar}
\left(\frac{1}{2} m \omega r_{i}^{2} + \frac{1}{\omega} C r_{i}x + C^{2} F[x,t]
\right)\right\} = G\left(r_{i}^{2}\right)
\exp\left\{-\frac{1}{\hbar}\left(\frac{1}{\omega} C r_{i}x + C^{2} F[x,t]\right)\right\},
\label{I2}
\end{equation}
where $C^{2} F[x,t]$ is the last term on the r.h.s. of (\ref{hdt}), and it does not depend on
the coordinate $r_{i}$ but it is a functional of $x(t)$ and we have the Gaussian
function $G( r_{i}^{2}) = \exp\left\{-\frac{m \omega r_{i}^{2}}{2\hbar}\right\}$. For the second
derivative $\frac{\partial^{2} \phi\left(t,x,r^{N}\right)}{\partial r_{i}^{2}}$, we get
\begin{equation}
\frac{\partial^{2} \phi\left(t,x,r^{N}\right)}{\partial r_{i}^{2}} = -\frac{m \omega}{\hbar} + \frac{1}{\hbar^{2}}m^{2} \omega^{2} r_{i}^{2} + \frac{1}{\hbar^{2}}\frac{C^{2} x^{2}}{\omega^{2}} + \frac{2C mX r_{i}}{\hbar^{2}}\,.
\label{I3}
\end{equation}
We decide to perform a calculation at the order $C^{2}$. All the quadratures can be performed
analytically since we have to deal with Gaussian integrals.
The final result is
\begin{equation}
I = - \frac{3}{4}\frac{C^{2} x^{2}}{\hbar m \omega^{2}}\phi(t,x).
\label{I4}
\end{equation}
The calculation of the second term
\begin{equation}
J = \frac{C x}{\hbar}\int \sum_{i}r_{i}\phi\left(x,r^{N},t\right)\rd r_{i}
\label{J1}
\end{equation}
can be performed exactly as for $I$ and at the second order in $C^{2}$ we find
\begin{equation}
\label{J2}
J = - \frac{C^{2} x^{2}}{\hbar m \omega^{2}}\phi(t,x).
\end{equation}
Finally the r.h.s. of (\ref{difnew}) is $(-I + J) = - \frac{1}{4}\frac{C^{2} x^{2}}{\hbar m \omega^{2}}\phi(t,x)$.

\ukrainianpart

\title{Часова еволюція невеликих систем з реакціями}
\author{Ж.-П. Бадіалі}
\address{LECIME, ENSCP-Університет П'єра і Марі Кюрі, CNRS/UMR7575,
75230 Париж, Франція}

\makeukrtitle
\begin{abstract}
Ми досліджуємо незворотну еволюцію невеликих систем, у яких
відбувається хімічна реакція. Ми ставимо подвійну мету: перша
вимагає знаходження рівняння, яке задає часово-зворотню поведінку,
друга полягає у побудові моделі з можливим точним розв'язком, щоб
зрозуміти основні події хімічної кінетики. Нашим головним знаряддям
є функція переходу, яка обчислює кількість шляхів, що з'єднують дві
точки у системі координат реакції. Розвинуто точне квантове рівняння
Смолуховського для системи з реакцією у вакуумі. У випадку
присутності розчинника, що перебуває у будь-який момент часу у
рівновазі із системою з реакцією, побудовано нове рівняння типу
Смолуховського. Обговорено перехід від квантової поведінки до
класичної. Також обговорено випадок системи з реакцією, яка не
перебуває у рівновазі з оточенням; він досліджується з використанням
інтегралів за траєкторіями і диференціальних рівнянь у частинних
похідних. Вивчено ефекти пам'яті та умови замикання. Для простої
моделі взаємодії точно  розрахована константа реакції, a також
обговорено, яке значення має енергія активації і фізичний зміст
множника перед експонентою.

\keywords незворотність, функція переходу, рівняння Смолуховського,
константа реакції, енергія активації

\end{abstract}


\begin{thebibliography}{99}
\bibitem{koslov}  Kosloff~R.,  J. Phys. Chem., 1988, \textbf{92}, 2087; \doi{10.1021/j100319a003}.
\bibitem{jpb1}  Badiali~J.P.,  J. Electroanal. Chem., 2011, \textbf{660}, 332; \doi{10.1016/j.jelechem.2011.07.016}.
\bibitem{jpb2}  Badiali~J.P.,  J. Electroanal. Chem., 2012, \textbf{676}, 40; \doi{10.1016/j.jelechem.2012.04.024}.
\bibitem{laidler}  Laidler~K.J.,   King~M.C.,  J. Phys. Chem., 1983, \textbf{87}, 2657; \doi{10.1021/j100238a002}.
\bibitem{eyring}  Eyring~H.,  J. Chem. Phys., 1935, \textbf{3}, 107; \doi{10.1063/1.1749604}.
\bibitem{trular1}  Garrett~B.C.,   Truhlar~D.G.,  J. Phys. Chem., 1979, \textbf{83}, 1052; \doi{10.1021/j100471a031}.
\bibitem{trular2}  Truhlar~D.G.,  Garrett~B.C.,   Klippenstein~S.J.,  J. Phys. Chem., 1996, \textbf{100}, 12771; \doi{10.1021/jp953748q}.
\bibitem{kramers}  Kramers~H.A.,  Physica (Utrecht VII), 1940, \textbf{4}, 284; \doi{10.1016/S0031-8914(40)90098-2}.
\bibitem{weiss}  Weiss~U.,  In: Quantum Dissipative Systems, World Scientific, Singapore, 2nd edition, 1999.
\bibitem{grabert}  Grabert~H.,   Schramm~P.,  Ingold~G.L.,  Phys. Rep., 1988, \textbf{168}, 115; \doi{10.1016/0370-1573(88)90023-3}.
\bibitem{caldeira}  Caldeira~A.O.,   Leggett~A.J.,  Physica~A, 1983, \textbf{121}, 587; \doi{10.1016/0378-4371(83)90013-4}.
\bibitem{ratner}  Nitzan~A.,   Ratner~M.A.,  Science, 2003, \textbf{300}, 1384; \doi{10.1126/science.1081572}.
\bibitem{jun} Jiang~J.,  Kula~M.,  Yi Luo, J. Chem. Phys., 2006, \textbf{124}, 034708; \doi{10.1063/1.2159490}.
\bibitem{galperin}  Galperin~M.,   Nitan~A.,  Ratner~M.A.,  Phys. Rev. B, 2007, \textbf{75}, 155312; \doi{10.1103/PhysRevB.75.155312}.
\bibitem{pecchia}  Pecchia~A.,   Romano~G.,   Di Carlo~A.,  Phys. Rev. B, 2007, \textbf{75}, 035401; \doi{10.1103/PhysRevB.75.035401}.
\bibitem{datta}  Datta~S.,  J. Phys. Condens. Mat.r, 1990, \textbf{2}, 8023; \doi{10.1088/0953-8984/2/40/004}.
\bibitem{sacha}  Kuznetsov~A.M., In: Stochastic and Dynamic Views of Chemical Reaction Kinetics in Solutions, Presses Polytechniques et universitaires romandes, Lausanne, 1999.
\bibitem{jpb4}  Badiali~J.P.,  Phys. Rev. E, 1999, \textbf{60}, 2533; \doi{10.1103/PhysRevE.60.2533}.
\bibitem{kampen} van Kampen~N.G.,  J. Stat. Phys., 1977, \textbf{17}, 71; \doi{10.1007/BF01268919}.
\bibitem{caldeira83}  Caldeira~A.O.,   Leggett~A.J.,  Ann. Phys., 1983, \textbf{149}, 374; \doi{10.1016/0003-4916(83)90202-6}.
\bibitem{feynman2}  Feynman~R.P.,  Hibbs~A.R., In: Quantum Mechanics and Path Integrals, chapter 3, p. 58, Mc Graw Hill, New York, 1965.
\bibitem{zwanzig}  Zwanzig~R.,  J. Chem. Phys., 1960, \textbf{33}, 1338; \doi{10.1063/1.1731409}.
\bibitem{jpb5}  Badiali~J.P.,  J. Phys. A Math. Gen., 2006, \textbf{39}, 7175; \doi{10.1088/0305-4470/39/23/001}.
\bibitem{jpb6}  Badiali~J.P.,  J. Phys. A Math. Gen., 2005, \textbf{38}, 2835; \doi{10.1088/0305-4470/38/13/002}.
\bibitem{jpb7}  Badiali~J.P.,  Condens. Matter Phys., 2000, \textbf{3}, 545; \doi{10.5488/CMP.3.3.545}.
\bibitem{jpb8}  Badiali~J.P.,  Preprint \arxiv{0902.0931}, 2009.
\bibitem{nagasawa}  Nagasawa~M., In Stochastic Processes in Quantum Physics, Monographs in Mathematics, 94, Birkhauser Verlag, Basel, 2000.



\end{thebibliography}
\end{document}